 \documentstyle[aps,preprint]{revtex}
\begin{document}
\preprint{Applied Physics Report 94-6}
\title
{Giant Conductance Oscillations in a Normal Mesoscopic Ring Induced by an SNS
Josephson Current}
\author{A. Kadigrobov$^{(1,2)}$\thanks{Permanent address: B.I. Verkin Institute
for  Low Temperature Physics and Engineering, 47 Lenin Ave., 310 164 Kharkov,
Ukraine},  A. Zagoskin$^{(1)}$\thanks{Email: alexz@fy.chalmers.se}, R.I.
Shekhter$^{(1)}$\thanks{Email: shekhter@fy.chalmers.se}, and M.
Jonson$^{(1)}$\thanks{Email: jonson@fy.chalmers.se}}
\address{Department
of Applied Physics$^{(1)}$ and Department of Physics$^{(2)}$ \\ Chalmers
University of Technology and G\"{o}teborg University, S-412 96 G\"{o}teborg,
Sweden}
\maketitle \draft

\begin{abstract}
A theoretical explanation of giant conductance oscillations observed in normal
mesoscopic rings with superconducting ``mirrors"  is proposed. The effect is
due
to resonant tuning of Andreev levels to the Fermi level, which enhances the
transparency of the system to the normal current. The mechanism is demonstrated
for a one-dimensional model system.
\end{abstract}

Recently, in a series of experiments by Petrashov et al. \cite{Petrashov}, new
unusual properties of mesoscopic silver rings in contact with superconducting
islands (``mirrors") were observed. These ``mirrors" could be placed across the
current leads (L-case) as in Fig.~1a, or as shown in Fig.~1b at the stubs
connected to the ring perpendicular to the normal current flow (T-case). The
three most striking features revealed in experiments are as follows:\\
(i) the amplitude of $hc/2e$-periodic Aharonov-Bohm (AB)
oscillations was at least 100 times larger in the L-case than in the ring
without  ``mirrors"; in the T-case the enhancement of $hc/2e$-oscillations was
about 10 times, and $hc/4e$-oscillations were observed as well; \\
(ii) in the
T-case the normal resistance of the system grows approximately twice, as the
superconductivity of the ``mirrors" is being suppressed
 by magnetic field (above 500 Gs);\\
(iii) the effects were totally absent in the case when one branch of the ring
was  completely crossed by a superconducting strip.

The confinement idea proposed in the original paper \cite{Petrashov}, i.e.
that the effect is due to confinement of quasiparticles in the ring by the
``mirrors", qualitatively explains the L-case behaviour, but  meets
difficulties both in  obtaining a proper magnitude of the effect in the L-case
(the maximum enhancement being of order 10), and in explaning the effect in
the T-case, where the quasiparticles are not confined in a longitudinal
direction.

An alternative explanation suggested by de Vegvar and Glazman \cite{deVegvar}
is based on a supposition that the  ``mirrors" induce superconductivity in a
significant portion of the ring itself. These superconducting parts play the
role
of filters, through which only Cooper pairs can pass. However, this mechanism
neither accounts for the significant difference between L- and
T-configurations,
nor for the complete absence of the effect when one branch of the ring was
crossed by a superconductor (as the authors of  Ref.~\onlinecite{deVegvar}
point out themselves).

In this paper we present some theoretical arguments which seem to explain the
physical nature of the observed phenomena; they are supported by a model
calculation. In our analysis we assume that the phase breaking length $L_{\phi}
= (D\tau_{\phi})^{1/2}$ and the normal metal coherence length $L_T = (\hbar
D/k_B T)^{1/2}$ are larger than the characteristic sample size $L$. Here
$\tau_{\phi}^{-1}$ is the inelastic scattering rate, $D$ is the diffusion
constant of quasiparticles, $k_B$ is the Boltzmann constant, and $T$ is
temperature. We will also assume that the ring is weakly coupled to the normal
reservoirs (source and sink of normal current).

We start from a qualitative discussion. If an electric
current is made to pass through a mesoscopic normal ring, it is well known
that its magnitude will oscillate as a function of the magnetic flux
threading the ring. These normal AB oscillations persist if the ring is brought
in contact with two superconducting mirrors. However, now a new group of
quasiparticles starts to contribute to the oscillations: the electrons (or
holes) that undergo Andreev reflections at the NS-boundary (between the ring
and one of the mirrors). We will see below that their contribution may dominate
the
AB oscillations under certain conditions.

In a ring which is  weakly connected to external reservoirs, the quantized
energy levels of quasiparticles are well defined \cite{Buttiker}. In our case
these levels are formed by reflections  both at NN-interfaces (between the ring
and the normal reservoirs) and at NS-boundaries with the ``mirrors". The energy
levels and are therefore sensitive to the magnetic flux threading the ring,
$\Phi$, {\em and} to the superconducting phase difference between the mirrors,
$\Delta\phi$. The corresponding quasiparticle states carry both the normal- and
the Josephson\cite{Buttiker,S} current.

Since the normal current in the linear response limit is carried only by the
quasiparticles on the Fermi surface, it is resonantly enhanced each time  an
energy level --- driven by the magnetic flux and/or the
superconducting phase difference --- passes the Fermi energy.
Simultaneously, the Josephson current changes sign.

 Therefore the relation can be guessed
\begin{equation} I_q \sim \frac{d I_J}{d \Delta\phi}. \label{2}
\end{equation}
As a matter of fact, using a 1D model (to be described below) we are able to
derive the relation valid in the ballistic case:
\begin{equation}
\lim_{\epsilon \rightarrow 0}\frac{I_q}{V \epsilon} = \left(  \frac{eL}{4\hbar
v_F}\frac{d I_J}{d \Delta\phi} +
 \frac{e^2}{\pi\hbar} \right). \label{4}
\end{equation}
Here $\epsilon$ is the probability for an electron to leave the ring for one of
the normal reservoirs through the NN-interface (see below).

 The second term in (\ref{4}) is due to the fact that the energy levels in the
ring are formed not only by Andreev reflections at NS-boundaries, but by the
normal reflections at NN-interfaces as well. Therefore their shift in a
magnetic field (and thus the conductance \cite{Kohn}) can not be completely
accounted for by an expression like (\ref{2}), where the magnetic flux, $\Phi$,
enters in a gauge invariant combination with the phase
difference between the superconducting ``mirrors", $\Delta\phi$. There must be
 a contribution to the level shift that depends solely on $\Phi$.

The relation (\ref{4}) shows that {\em if} there is a Josephson current in the
system, it leads to a resonant contribution to the normal current for certain
values of the magnetic flux. Indeed, the phase dependence of the Josephson
current in an SNS contact has a sawtooth-like form, which would give rize to
$\delta$-function shaped peaks when its derivative is taken with respect
to phase \cite{S}. The question is, how can a stationary supercurrent
flow between two finite and small superconducting ``mirrors"?

Let us first consider a standard SNS-junction in an external  magnetic field,
{\bf B},
parallel to the boundary between the normal layer and superconducting
half-spaces (Fig. \ref{f.1}a). The phase difference  between the
superconductors
is \cite{S}
\begin{equation}
\Delta\phi(y) = \phi_0 - \frac{2eBLy}{\hbar c} \equiv \phi_0 -
2\pi\frac{2\Phi(y)}{\Phi_0},
\label{5}
\end{equation}
where $y$ is the coordinate on an axis that lies in the plane of the normal
layer
and is normal to the magnetic field, $\phi_0$ is an arbitrary phase difference
between two isolated superconductors, and $L$ is the width of the  normal
layer.
The quantity $\Phi(y_0)$ is evidently the magnetic flux through the part of the
junction between the lines $y=0$ and $y=y_0$, and $\Phi_0 = hc/e$. The
distribution of Josephson current is schematically shown in Fig.\ref{f.1}a;
each
current line encircles one half flux quantum (Josephson vertex).

In a real experimental situation the normal stubs connecting the metal ring
 with the ``mirrors" have a finite width. This creates the situation when equal
and opposite Josephson currents flow along the  edges of the stubs (see
Fig.\ref{f.1}b).   In order to establish such a current distribution,  the
phase difference between the superconducting ``mirrors" tunes to each given
value of the
penetrating magnetic flux  (via changing the constant
$\phi_0$ in (\ref{5})). Therefore the partial Josephson currents  will
oscillate,  contributing
to  the AB oscillations in the normal conductance of the ring, though their sum
(net Josephson current between the ``mirrors") is always zero.

Now we can discuss when this contribution  can  play a major role. It is known
that the amplitude of the Josephson current density is not sensitive to
an increase in the area of an SNS junction \cite{S}. On the other hand, the
amplitude of  the conductance oscillations in a normal metal ring decreases by
a factor $\sqrt{N_{\perp}}$ as its
cross-section grows \cite{Imry} [the number of channels in a ring
with the cross-section area $A$ is $ N_{\perp} \sim A/\lambda_F^2]$. The reason
is that in the former case it is the
effective momentum of the electron-hole excitations that carry the Josephson
current, which is quantized:
\begin{equation} p_{\parallel}^{(e)} -
p_{\parallel}^{(h)} = \sqrt{ 2m(E_F + E) - p_{\perp}^2} -  \sqrt{ 2m(E_F - E) -
p_{\perp}^2} \simeq \frac{2mE}{\sqrt{E_F - p_{\perp}^2}}, \label{SNS}
\end{equation}
while in the latter case it is the electronic momentum itself:
\begin{equation}
p_{\parallel} \simeq \sqrt{ 2mE_F  - p_{\perp}^2} +
\frac{mE}{\sqrt{ 2mE_F  - p_{\perp}^2}}. \label{N}
\end{equation}
The first term
in (\ref{N}), which is absent in (\ref{SNS}), causes fast oscillations when the
partial current is integrated over the transverse momentum $p_{\perp}$. This is
the reason of drastical  reduction of the ``normal" AB oscillations' amplitude
in a ring with large cross-section.  In this case the ``Josephson" contribution
will be dominating.

The above considerations lead us to the conclusion that the ratio of the
amplitude of AB oscillations  in a ring with superconducting ``mirrors" to
the amplitude of AB oscillationsits in a ``normal" ring is  of the order of
$\sqrt{N_{\perp}}$.

Let us make some numerical estimates. Taking for the cross-section area the
value
$A \approx 5 \cdot 10^{-11}
{\rm cm}^2$, qhich is consistent with the experiment \cite{Petrashov}, and
$\lambda_F \sim 10^{-8} {\rm cm}$, we find that the enhancement of the AB
conductance oscillations due to the proposed mechanism is $\simeq 700$ times
compared to the ring without superconducting ``mirrors". (The actual value of
Ref.\cite{Petrashov} was up to 400 times).

The above qualitative speculations can be corroborated by a model calculation.
The model system is shown in Fig.\ref{f.3}. It contains all the physically
significant features of the experimental setup (Fig.\ref{f.0}); the normal
%
%
part of the current from reservoir $I$ to reservoir $II$ through leads 3, 1,
and 4 is controlled by the Josephson current between the ``mirrors" through
leads 1 and 2.

All the normal scattering in our model is confined to the nodes $A$ and $B$.
These  are described by identical $6 \times 6$ S-matrices , which relate the
incoming and outgoing wave amplitudes of the quasiparticles in the 1D wires. We
use real matrices parametrized by a real number $0 \leq \epsilon \leq 1/2$
\cite{Buttiker}, which in the limit $\epsilon \ll 1$ (weak coupling to the
reservoirs, i.e. small transition probability from lead 1 to   leads 3 and 4)
have the form
\begin{equation}
    {\bf S} = \left(
     \begin{array}{ccc}
               -\epsilon/2 \cdot\hat{\bf 1}   &  (1-\epsilon/2)
\cdot\hat{\bf 1}  &  \sqrt{\epsilon} \cdot\hat{\bf 1} \\
                 (1-\epsilon/2) \cdot\hat{\bf 1}  &    -\epsilon/2
\cdot\hat{\bf 1}     &  \sqrt{\epsilon} \cdot\hat{\bf 1}    \\
               \sqrt{\epsilon} \cdot\hat{\bf 1}     &   \sqrt{\epsilon}
\cdot\hat{\bf 1}   &  (-1+\epsilon) \cdot\hat{\bf 1}
\end{array}                          \right). \label{29} \end{equation}
Here $\hat{\bf 1}$ is the unit $2 \times 2$-matrix, which reflects the fact
that in the presense of an  NS-boundary we have to use the two-component  wave
function of the quasiparticle even in the normal leads, in order to account for
the electron-hole correlations created by Andreev reflections \cite{S}.
Eq.(\ref{29}) reflects the fact that the electron- and holelike excitations
are not mixed in the  nodes {\em A} and $B$. On the other hand, we assume that
at
 the interfaces of leads 1, 2 and the
superconducting islands only   Andreev   reflection  takes place,   and the
 normal reflection is absent.

The calculations to be described are straightforward. We are interested in the
linear response value of the normal conductance. Therefore,  the problem is
reduced to the calculation of the transition probability of a quasiparticle
from, say, reservoir $I$ to reservoir $II$.   The initial scattering matrices
can be replaced by effective matrices of dimensionality $4 \times 4$, ${\cal
S}_A$ and  ${\cal S}_B$. They  only relate the quasiparticles in leads 3, 4,
and the portion of lead 1 between A and B to each other and include the effects
of Andreev reflections (see \cite{K2} for details). We have a standard
Landauer configuration with two reservoirs connected by a 1D wire with
scatterers A and B, but with two-component wave functions of the
quasiparticles;
 the components are mixed by Andreev reflections at NS interfaces.  The
conductance is  then obtained as \cite{Lambert}
\begin{equation}
G = \frac{2e^2}{h}\cdot 2 \int_{0}^{\infty} d\xi  \left(  T_0^> +  R_a^>
\right)
\left(-\frac{\partial n_F(\xi)}{\partial \xi}\right) + \eta, \label{G}
\end{equation}
where $T_0^> \:\: ( R_a^> )$  is  the normal transition (Andreev reflection)
probability for an electron incident from the left   normal reservoir;
$n_F(\xi)$ is the Fermi distribution, and $\xi$ is energy measured from the
Fermi level. The term $\eta$ in (\ref{G})  quickly oscillates as a function of
the electron momentum (as $\sim \exp 2p_F L$, $L$ being the length of lead 1)
\cite{K2} and  is exactly zero in the case of time reversal symmetry
\cite{Lambert}.

After averaging over fast spatial oscillations on the scale of $\lambda_F \ll
L$,   the  coefficients in (\ref{G}) take the following
values \cite{K2}:
\begin{equation}
T_0^>(\xi) \approx R_a^>(\xi) \approx  \epsilon^2 \left(
\left|a_+(\xi)\right|^{-2} +
\left|a_-(\xi)\right|^{-2} \right).
\end{equation}
The resonant denominators $\left|a_{\pm}(\xi)\right|^2$ vanish close to
the energies $\xi_n^{\pm}$ of the Andreev levels in lead 1:
\begin{eqnarray}
\left|a_{\pm}(\xi)\right|^{-2} \approx \sum_{n} \left\{ \left(\frac{2L}{\hbar
v_F}\right)^2 \cdot
\left( \left(\xi - \xi_n^{\pm}\right)^2 + \epsilon^2
\left(\frac{\hbar v_F}{4L}\right)^2 \right) \right\}^{-1}; \label{a+-} \\
\xi_n^{\pm} =  \frac{\pi\hbar v_F n}{L} + \frac{\hbar v_F}{2L}
(\pi \mp 2\pi\frac{2\Phi}{\Phi_0}).
\label{levels}
\end{eqnarray}
Here $v_F$ is the Fermi velocity and $\Phi_0 = hc/e$ is the magnetic flux
quantum.  In place of the superconducting phase difference, $\Delta\phi$, only
the  (dimensionless)  magnetic flux $\Phi/\Phi_0$ through the loop formed by
the normal wires 1, 2 and the superconducting mirrors enters expression
(\ref{levels}) for the Andreev energies. This is due to the fact that the
Josephson current in lead 1 must be exactly cancelled by the one in lead
2. This condition fixes the phase difference between the
superconducting mirrors.

Provided that the Andreev level separation  exceeds the level width
$\Delta E = \epsilon \hbar v_F/ 4L$, a condition which is satisfied for small
enough $\epsilon$, we can calculate the integral  in  (\ref{G}) using the
Poisson summation formula. At zero temperature one finds the expression
\begin{equation}
G = \epsilon \frac{e^2}{h} \left( 1 + 2 \sum_{n=1}^{\infty}
(-1)^n e^{-2|n|\epsilon} \cos \left(n \cdot
2\pi\frac{2\Phi}{\Phi_0}\right)\right).  \label{GGG}
\end{equation}
If we
compare this result to the well known expression for the Josephson current in a
planar SNS-junction \cite{S},
\begin{equation}
I_J (\Delta\phi)= \frac{8 e
v_F}{\pi L} \sum_{n=1}^{\infty} \frac{(-1)^{n+1}  \sin n\Delta\phi}{n},
\end{equation}
we see that indeed, in the weak coupling limit ($\epsilon
\rightarrow 0$) we have the relation between the normal conductance and the
Josephson current given by equation (\ref{4}).

The above calculations are directly generalized to the case of $N_{\perp} > 1$
transverse modes in the normal wire, provided that they are not mixed  by
scattering.  Then the conductance (\ref{GGG}) should be simply multiplied by
$N_{\perp}$.
The fact that the amplitude of its oscillations now can exceed the conductance
quantum, $2e^2/h$, reflects the ballistic character of the system under
consideration.

In conclusion, we have demonstrated that the normal conductivity of a
mesoscopic ring with superconducting ``mirrors" is  sensitive to the Josephson
current between them.  The corresponding contribution to the conductance is
resonant and in a many-channel case  gives rize to greatly enhanced
Aharonov-Bohm conductivity oscillations in the system. The results provide an
explanation to recent experimental results by Petrashov et al.
\cite{Petrashov}.

\acknowledgements

The authors are grateful to V. Antonov, C.W.J. Beenakker, T. Claeson, P.
Delsing, Yu. Galperin, L. Gorelik,  V. Petrashov, S. Rashkeev, V. Shumejko, and
A. Slutskin for many fruitful discussions.

This work was supported by the Royal Swedish Academy of Sciences (KVA), the
Swedish Natural Science Research Council (NFR), and by  the Swedish Board
for Industrial and Technical Development (NUTEK). One of us (A.K.) gratefully
acknowledges the hospitality of the Department of Applied Physics and
Department of Physics, Chalmers University of Technology and G\"{o}teborg
University.

\begin{figure}
\caption{Sketch of the experimental setup of  a mesoscopic ring
with superconducting  ``mirrors" used in Ref.~[1]. {\em I, II}
are normal reservoirs, S - ``mirrors" (small superconducting islands). The
magnetic field is normal to the picture plane. (a) L-configuration. (b)
T-configuration.} \label{f.0}  \end{figure}

\begin{figure}
\caption{(a) Distribution of Josephson current in an SNS-junction in the
presence of a magnetic  field; (b) Josephson current in a mesoscopic ring with
superconducting ``mirrors"} \label{f.1}
\end{figure}

\begin{figure}
\caption{Model system where 1, 2, 3, and 4 are ideal normal-conducting 1D
leads.
Nodes  {\em A, B} are described by real scattering matrices (see text). The
directions of normal ($I_q$) and Josephson ($I_J$) currents are schematically
shown. The magnetic field {\bf B} is normal to the plane of the picture,
and the distance between the ``mirrors" equals $L$.} \label{f.3}
\end{figure}

\end{document}